# First numerical analysis of runaway electron generation in tungsten-rich plasmas towards ITER

J. Walkowiak[1,2,*], M. Hoppe[3], I. Ekmark[4], A. Jardin[1], J. Bielecki[1], K. Król[1], Y. Savoye-Peysson[5], D. Mazon[5], D. Dworak[1] and M. Scholz[1]

[1] Institute of Nuclear Physics Polish Academy of Sciences, PL-31342 Krakow, Poland
[2] National Centre for Nuclear Research (NCBJ), 7 Andrzeja Sołtana Str., Otwock 05-400, Poland
[3] Fusion Plasma Physics, Department of Electric Energy Engineering, KTH, 10044 Stockholm, Sweden
[4] Department of Physics, Chalmers University of Technology, Gothenburg SE-41296, Sweden
[5] CEA, IRFM, F-13108 Saint-Paul-lez-Durance, France

E-mail: jedrzej.walkowiak@ifj.edu.pl

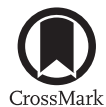

## Abstract
The disruption and runaway electron analysis model code was extended to include tungsten impurities in disruption simulations with the aim of studying the runaway electron (RE) generation. This study investigates RE current sensitivity on the following plasma parameters and modelling choices: tungsten concentration, magnetic perturbation strength, electron modelling, thermal quench time and tokamak geometry—ITER-like or ASDEX-like. Our investigation shows that a tungsten concentration below $10^{-3}$ does not cause significant RE generation on its own. However, at higher concentrations it is possible to reach a very high RE current. Out of the two tested models of electrons in plasma: fluid and isotropic (kinetic), results from the fluid model are more conservative, which is useful when it comes to safety analysis. However, these results are overly pessimistic when compared to the isotropic model, which is based on a more reliable approach. Our results also show that the hot-tail RE generation mechanism is dominant as a primary source of RE in tungsten induced disruptions, usually providing orders of magnitude higher RE seed than Dreicer generation. We discuss best practices for simulations with tungsten-rich plasma, present the dependence of the safety limits on modelling choices and highlight the biggest shortcoming of the current simulation techniques. The obtained results pave the way for a wider analysis of tungsten impact on the disruption dynamics, including the mitigation techniques for ITER in the case of strong contamination of the plasma with tungsten.

Keywords: runaway electrons, tungsten impurities, ITER, computational plasma physics

(Some figures may appear in colour only in the online journal)

---

* Author to whom any correspondence should be addressed.

## 1. Motivation

Runaway electrons (RE) are one of the major issues for tokamak safety. When unmitigated, they can cause local melting of the plasma-facing components (PFC) and sometimes even damage the underlying components of the machine [1, 2]. Extensive work has been done to develop techniques for suppressing RE generation during the disruption, e.g. by massive material injection [3]. Unfortunately, active methods of mitigation require the disruption to be predicted in time and a finely tuned material injection, to avoid a too fast current quench (CQ), which can lead to mechanical failure of the machine components.

Some disruptions occur very suddenly, without the usual signals of growing plasma instability that allow the triggering of the mitigation system. One such scenario is the impurity influx resulting from large sources such as flakes or dust, which is often labelled as UFO (as such particles were described as unidentified flying objects), or unidentified impurity influx [4–6]. These events are responsible for up to a few percent of the disruptions in present-day tokamaks, but their occurrence is strongly related to the PFC condition. In some campaigns, UFOs are practically not occurring, while if the machine experienced some significant damage such as PFC melting, they sometimes become a significant issue [7]. In reactor-scale tokamaks, with discharge times extended into minutes, PFC degradation can occur even as a result of normal operation conditions [8]. The impact of PFC degradation on the impurity influx into the plasma in the long-term operation is still an open question.

The effects of UFOs can vary depending on the PFC composition. Operation at JET showed that carbon wall tokamaks cannot serve as thermonuclear reactors due to their tritium retention [9]. This resulted in increased use of tungsten (W) as plasma-facing material [8, 10, 11]. Unfortunately, W is a heavy element with $Z = 74$, which creates significant problems when entering the plasma as an impurity. Complete ionisation of $W$ requires an energy of $\sim$80 keV [12], much higher than plasma temperatures foreseen even in the largest tokamaks such as ITER or DEMO [13]. This results in strong line radiation of partially ionised W impurities, orders of magnitude higher than for low-$Z$ elements of similar concentration. Therefore, the $W$ impurity concentration in the plasma must be kept at a much lower level than in the case of carbon impurities. Concentrations limited to a relatively small fraction $10^{-5}$–$10^{-4}$ are required for a positive energy balance in a fusion reactor [14].

When the $W$ concentration rises in the plasma core, it can break the H-mode or lead to a disruption. It is worth mentioning that $W$ influx can lead both to slow accumulation in the plasma core, which at some point causes an MHD instability, or to much faster impurity influx observed as a radiation spike [15]. In the former case, it is usually possible to mitigate a disruption, but the latter case can be much more problematic. A characteristic for high-$Z$ impurities is that sometimes the thermal quench (TQ) does not lead to a CQ and is instead followed by at least partial temperature recovery and a very slow CQ, or even recovery close to pre-disruptive conditions [5, 15]. This can be the case when tungsten impurities are expelled from the plasma by a MHD instability [16, 17]. However, it seems that producing such disruptions on purpose in a reliable way is beyond control capabilities.

With plans to start ITER with tungsten PFCs, it is necessary to estimate the risk posed by tungsten impurities to every aspect of machine operation and safety. For REs, tungsten was not yet included in the analysis and this is the gap which this work aims to address. It was done in a series of numerical experiments conducted with the disruption and runaway electron analysis model (DREAM) code [18], collecting results from a few hundreds of plasma simulations in total. DREAM, for the purpose of this work, has been extended with the necessary atomic data, to include $W$ impurities in the simulations. The basic assumptions of the simulations and their limitations are introduced in section 2. Section 3 describes the investigated simulation approaches and their consequences. It is divided into the following parts:

- Comparison of two models of the electron distribution: fluid and kinetic.
- Impact of magnetic perturbation strength during TQ and CQ on the RE generation. In DREAM, stochastization of the magnetic field lines during disruptions is modelled by an effective radial diffusion, because calculation of the magnetic flux surfaces would be very expensive and sensitive to initial conditions [19, 20].
- The contribution of the Dreicer mechanism to overall RE generation and different approaches to the Dreicer generation modelling with a fluid plasma model (Connor–Hastie and neural network).
- Various approaches to the TQ time definition and effects of varying the TQ period in simulations, when radial transport is enhanced.
- Influence of the device parameters on the RE generation in a W induced disruption.

Section 4 describes the results in an ITER-like disruption scenario, which is followed by a discussion of the consequences for the ITER disruption mitigation system analysis.

## 2. Methods

Results presented in the following sections were obtained with the DREAM code, which allows for self-consistent simulation of plasma cooling and associated RE dynamics during disruptions. The code can fully-implicitly solve a set of nonlinear coupled equations describing the evolution of temperature, density, current density and electric field, as well as the full electron distribution function in arbitrary axisymmetric geometry. It employs a combination of fluid models for background plasma parameters, including the toroidal electric field, electron and ion temperatures, ion densities and charge states, as well as various models for REs, ranging from fluid to fully kinetic. The most complete model included in DREAM is drift-kinetic model with a fully relativistic Fokker–Planck test-particle operator for electron-electron collisions, synchrotron radiation reaction force, an avalanche operator, bremsstrahlung and screening effects in a partially ionised plasma. The field-particle part of the collision operator is neglected, which would result in underestimated conductivity. To amend that, the ohmic current is corrected with a conductivity correction, to capture the correct Spitzer response to an electric field [18]. In the presented work, we used a feature of DREAM which allows parts of the electron phase space to be modelled kinetically, and the remainder to be described by fluid equations. The two approaches were compared, one with a fully fluid representation of plasma, and the other one with a reduced kinetic model, where suprathermal electrons are modelled with kinetic equations and the bulk of electrons is modelled with a fluid approach. A detailed description of differences between these two models is presented in section 3.3.

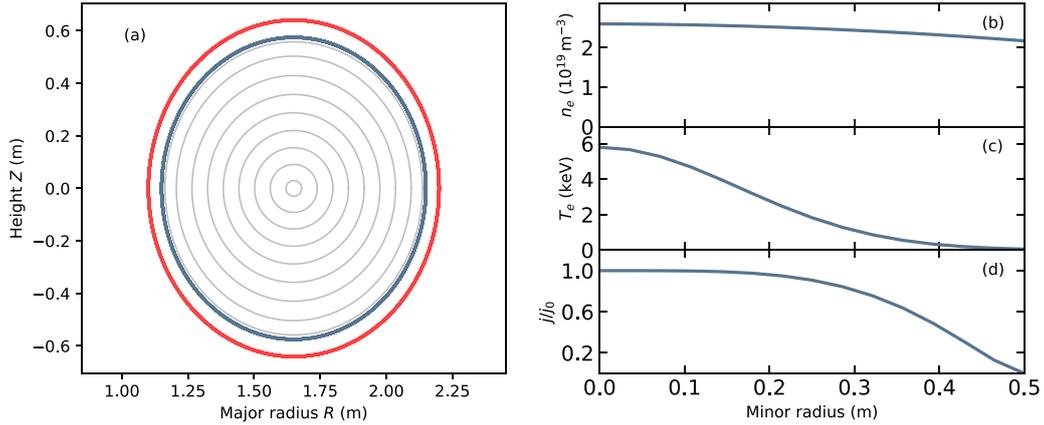

**Figure 1.** Parameters for the baseline ASDEX-like scenario: (*a*) magnetic field flux surfaces (gray), with the plasma boundary shown in blue and vessel wall in red. Profiles of: (*b*) initial electron density $n_e$, (*c*) initial electron temperature $T_e$, and (*d*) initial plasma current density $j/j_0$, $j_0 = 1.62 MA\,\text{m}^{-2}$ [18]. Reproduced from [18]. CC BY 4.0.

Each disruption simulation was divided into two phases: TQ and CQ. The first phase is a short period at the beginning of the disruption, when the temperature drops rapidly. During the TQ, the magnetic flux surfaces are usually destroyed by intense MHD activity in a process called stochastization of the magnetic flux surfaces [19]. However, this is a very complex process, a detailed simulation of which would be time consuming and possibly too sensitive to initial conditions to make it useful in our study. Instead, elevated transport coming from destruction of magnetic flux surfaces is modelled as increased diffusion of heat [21] and RE current [22]. While newer transport models have been developed, they are either not applicable in this scenario, or would require input from 3D MHD simulations. Furthermore, due to the way how these operators are implemented, using a more recent transport model would effectively correspond to using a slightly different $\delta B/B_0$ value. Since $\delta B/B_0$ is not determined self-consistently, but is one of the input parameters whose impact is investigated in this work, the choice of exact transport model should not change the final conclusion. During the CQ, RE diffusion is set to zero while a weak magnetic perturbation is prescribed for the heat diffusion in order to maintain a small amount of heat transport and avoid unphysical effects [23]. For numerical stability, change of the transport parameters is not instantaneous, but takes place over the last 1% of the TQ time. Otherwise, the CQ simulation would be prone to diverge at the first time step. The moment at which the disruption transfers from the TQ to the CQ phase is difficult to define, so this point is investigated in detail in section 3.1.

### 2.1. Tokamak geometry and plasma parameters

Simulations were conducted with two scenarios: the baseline discharge presented in the DREAM reference paper [18], which was created to resemble experimental results from ASDEX-U, and the ITER-like scenario modelling a disruption during the ITER H-mode flattop phase. Radial profiles of plasma electron density $n_e$, electron temperature $T_e$ and normalised plasma current density $j$ are presented in figures 1 and 2 for ASDEX-like and ITER-like cases, respectively. Geometry and plasma parameters are listed in table 1. The two sources of primary RE are Dreicer and hot-tail generation. RE generation by tritium decay and Compton scattering are neglected as they do not originate from the presence of impurities in the plasma and are relevant mainly for burning plasma scenarios. Avalanche RE generation is included.

In theory it is possible for gamma radiation to be present due to bremsstrahlung interaction between REs and tungsten impurities, but at the given plasma density the total fraction of bremsstrahlung reactions leading to the creation of another RE is negligible. Because of very small scattering cross-section, plasma can be in this matter treaded as optically thin, meaning that almost all photons created in the plasma will leave its volume without interactions with other particles. Compton scattering is important source when gamma radiation comes from external sources like walls, because it generates primary RE. When radiation is created by interaction

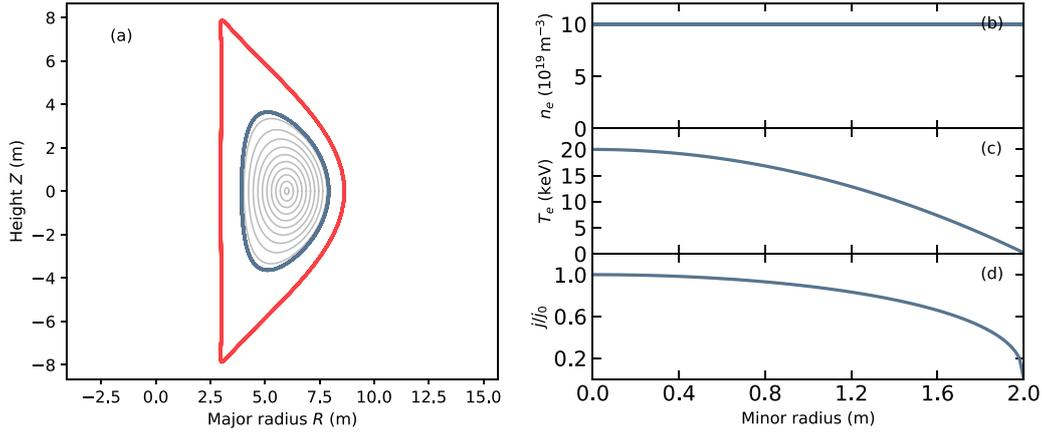

**Figure 2.** Parameters for the baseline ITER-like scenario: (*a*) magnetic field flux surfaces (gray), with the plasma boundary shown in blue and vessel wall in red. Profiles of: (*b*) initial electron density $n_e$, (*c*) initial electron temperature $T_e$, and (*d*) initial plasma current density $j/j_0$, $j_0 = 1.12\, MA\, m^{-2}$ [24]. Reproduced from [23]. CC BY 4.0.

**Table 1.** Tokamak geometry and plasma parameters of the ASDEX-like and ITER-like scenarios Adapted from [18]. CC BY 4.0.

| Parameter | ASDEX-like [18] | ITER-like [24] |
| --- | --- | --- |
| Major radius $R_m$ | 1.65 m | 6.0 m |
| Minor radius $a$ | 0.5 m | 2.0 m |
| Wall radius $b$ | 0.55 m | 2.833 m |
| Elongation at edge $\kappa(a)$ | 1.15 | 1.82[a] |
| Toroidal magnetic field $B_0$ | 2.5 T | 5.3 T |
| Initial plasma current $I_{p,0}$ | 800 kA | 15 MA |
| Resistive wall time | 10 ms | 500 ms |

[a] Elongation in the ITER-like example is not homogeneous, but varies in the range 1.5–1.82 as the elongation is described with $\kappa = 1.5 + 0.02 r^4$, where $r$ is the plasma minor radius coordinate.

of RE, it would be only small correction to the avalanche term.

During a disruption, part of the current is induced in the tokamak wall. The effective wall conductivity is included by prescribing the resistive wall time ($\tau_W = L_W/R_W$), whose exact value can be obtained from detailed simulation of the vacuum vessel inductance $L_W$ and resistance $R_W$ or measurements on existing devices. Neglecting the wall conductivity can lead to unphysical results, as it affects the disruption dynamics and current distribution.

### 2.2. Tungsten impurities

Tungsten impurities are introduced at the beginning of the simulation, as neutral atoms with an initial temperature of 1 eV. The *W* concentration is defined as $n_W/n_{e,0}$, where $n_W$ is W density and $n_{e,0}$ is the initial density of free electrons. In order to identify the main factors which influence RE generation in tungsten-rich plasmas, $n_W$ is assumed to be spatially uniform and constant during the whole simulation. Assuming a strongly peaked impurity profile in the simulation would result in local plasma cooling and subsequent current redistribution. In reality, this results in an MHD instability, followed by magnetic field line stochastization and elevated transport. When magnetic surfaces are destroyed, the impurities should be diffused across plasma which would result in a mostly flat W concentration profile. Reproducing such effects in detail would require significant computational effort and is outside the scope of this work. When a constant, strongly inhomogeneous *W* profile is prescribed in the simulation, there is no CQ, as part of the plasma is not sufficiently cooled and some current can be maintained by redistribution to the hot region. The assumption of a flat tungsten concentration profile simplifies the interpretation of the results and should not be far from reality shortly after the TQ.

Performing simulations with *W* impurities in DREAM required some additional atomic data, which has not been included earlier in the code. Most of the necessary quantities were obtained from the ADAS database [25]. To calculate inelastic collisions in the Fokker–Planck collision operator, the Bethe stopping power theory is used [26]. This requires the mean excitation energy (MEE) for tungsten ions, which is not available neither from measurements nor from present ab-initio theoretical calculations. MEE was calculated with an approach based on the local plasma approximation (LPA) corrected to fit the results from Sauer *et al* [27–29]. More detailed comments on the selection of ADAS coefficients and MEE calculation can be found in appendices A and B.

### 2.3. Magnetic perturbation

One of the main parameters which have a strong influence on the disruption dynamics is the magnetic perturbation due to magnetic flux surface stochastization during the TQ. Nevertheless, the exact simulation of the disruption with magnetic surfaces is usually not feasible due to the large number of unknowns in the plasma initial state and exact disruption evolution, which would require much more detailed simulations at extremely high computational costs. Therefore, averaged magnetic perturbation strengths were provided instead, which were given with respect to the magnetic field strength on the magnetic axis and denoted as $\delta B/B_0$. Based on the numerical simulations of MHD instabilities, it is expected to be in the range of $10^{-3}$–$10^{-2}$ [23, 30]. As diffusion coefficients in the simulation depend on this parameter, it has a direct effect on the temperature evolution and RE confinement in the initial phase of the disruption. As long as the *W* concentration is not extremely high, it will have an effect on the proportion of the energy lost by radiation to the energy lost through the diffusion to the walls.

We conducted a parameter scan with different tungsten concentrations and $\delta B/B_0$ values for the TQ. The W concentration was varied from $10^{-3}$ to $3.16 \times 10^{-1}$. At lower concentrations, tungsten has a very weak effect on plasma dynamics. The upper limit was selected arbitrarily, as reaching such high concentrations is rather improbable in tokamaks, nevertheless it is useful to show the trends with rising *W* concentration. $\delta B/B_0$ was varied between $10^{-3}$ and $10^{-2}$. Values for both the *W* concentration and magnetic perturbation were distributed logarithmically with two values per order of magnitude, giving 6 values of *W* concentration and 3 values of $\delta B/B_0$. During the CQ, the heat diffusion was set with $\delta B/B_0 = 4 \cdot 10^{-4}$, to avoid unphysical concentration of current into hot channels. This value was used before in other works [24, 31] and we confirmed in test simulations that also in our case lower values can lead to such unphysical effects. The transition between TQ and CQ, when the transport coefficients are changed according to decreasing $\delta B/B_0$, takes place during the last 1% of the TQ time. As the real duration of this transition is impossible to accurately predict within the presented approach, we used a transition time which was as short as possible from the numerical stability point of view.

The scan was repeated for different simulation approaches and the results are presented in the following section. Unfortunately, for some simulations the solver failed to find a convergent solution. The failure to converge is usually related to the Newton solver and not to the given set of physical equations. This is a common numerical issue in many computational tools, usually related to the too large time step, which leads to overshoots of predictions when the time derivative is large. In some cases the reason is different and usually difficult to identify. In our analysis the convergence was usually ensured by a change of the time step. In cases where we could not obtain convergence in any known way, we do not present the results.

## 3. Investigated simulation approaches

### 3.1. TQ time

TQ is the initial period during the disruption, when plasma temperature decreases rapidly. It is related to increased transport coefficients due to strong magnetic perturbation when the magnetic surfaces are broken. In the simulation TQ refers to the period when diffusion coefficients are increased, which results in much faster temperature decrease than in the CQ. Despite the fact that in simulation the temperature decrease is just the result of the TQ, not the cause, we use it to define the TQ duration. In general, there can be disruptions with incomplete TQ, where the magnetic perturbation stops before a very low temperature is reached [15]. However, we focus on the most pessimistic scenario of a disruption with a single strong TQ. In mitigation scenarios with low-Z impurities like neon or argon, there is a rapid increase of the impurity cooling factor when the plasma temperature drops below 100 eV as nearly stripped impurity ions recombine [14]. The radiated power can then rise by two orders of magnitude. This makes the cooling of the plasma very rapid below 100 eV, so the TQ time can be straightforwardly defined as the time needed to reach temperatures of a few tens eV.

However, *W* impurities radiate strongly in temperatures up to few keV—this is an effect of much higher atomic number of tungsten ions. Because of that, there is no steep radiation increase at low temperatures. Defining a TQ time is then more difficult, because the temperature drop will stop when the radiated power is balanced by ohmic heating without necessarily reaching the prescribed temperature. As a first approach, we used a definition of the TQ time ($\tau_{TQ}$) similar to the definition of the CQ time [23]. It is formulated as follows:

$$\tau_{TQ} = \frac{t(T = 0.2T_0) - t(T = 0.8T_0)}{0.6} \quad (1)$$

where $T_0$ is the initial temperature before the disruption.

In this approach, hereafter referred to as 80–20, the TQ time is defined as the time it would take for the temperature to drop from its initial value $T_0$ to 0, assuming the temperature drops at the same rate as from 80% to 20% of $T_0$.

In general, many other definitions can be used, for example by changing time measurement points to 90% and 10% of the initial temperature. We decided to investigate the effects of our TQ time definition by comparing it with a different approach, which provided a longer TQ time and was based on the measurement of the temperature time derivative, d$T$/d$t$. This definition is based on the assumption that the time derivative becomes very small at the end of the TQ. The problem is that the magnitude of the time derivative of the temperature can be very different depending on the magnetic perturbation strength and amount of impurities, which define the energy losses. To solve this problem, we calculate d$T$/d$t$ and normalise it to its maximum value, which is reached in the first time steps of the simulation. We assumed that when normalised d$T$/d$t < 10^{-4}$, then the TQ is over. While the exact number is selected arbitrarily, it is in the order of magnitude which provides a TQ time a few times longer than the 80–20 definition, while still being robust enough to be useful in most of the simulated cases.

As shown in figure 3, the TQ time can have an impact on the initial seed of REs for the avalanche occurring during the CQ. The increased radial diffusion accelerates the temperature drop, which is a driving force for hot-tail generation. But at the same time, REs are deconfined, so the overall RE population will peak and then decline if the TQ is long enough. This is in line with our expectations, as intentionally applied magnetics perturbations of sufficient strength were already recognized as one of the possible ways to mitigate RE [32]. The exact TQ time will be decisive for the initial RE population for the CQ and thus will influence the final RE current during the plateau phase. The dependence of the RE current at the moment of transition from TQ to CQ is presented on figure 4. The 80–20 definition seems to be a safe choice for the TQ time definition. Indeed, it is fast enough to break the TQ before REs are almost fully deconfined, so there is no risk that the initial RE population of the CQ will be strongly underestimated. It seems there is no need to catch the maximum of the RE population during the TQ. From the point of view of breaching the safety limits, it can be important only for cases at the border of allowed RE current, but in such cases the RE generation is very weak even during the TQ and will be negligible, if any during the CQ. The difference in the RE current generated in both approaches is presented in figure 5.

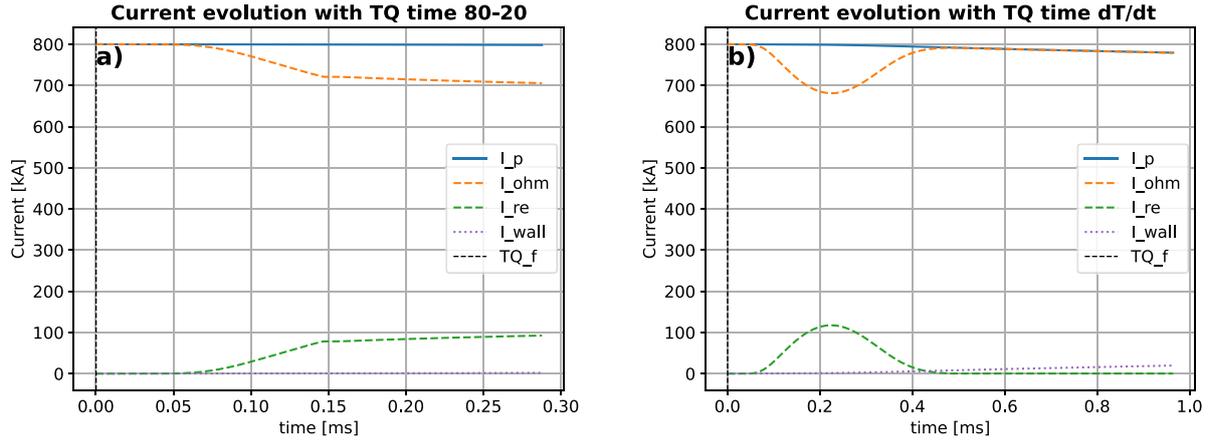

**Figure 3.** Time evolution of total plasma current (I_p), ohmic current (I_ohm), RE current (I_re), wall current (I_wall). Transition from TQ to CQ is marked with a dashed black line (TQ_f). Simulation was performed with ASDEX-like parameters and fluid model. W concentration was $1\times10^{-2}$, $\delta B/B_0$ in TQ was $3.16\times10^{-3}$, $\delta B/B_0$ in CQ was $4\times10^{-4}$. (*a*) Results from the simulation with TQ time obtained from the 80–20 definition. (*b*) Results from simulation with the TQ time obtained from the d$T$/d$t$ definition.

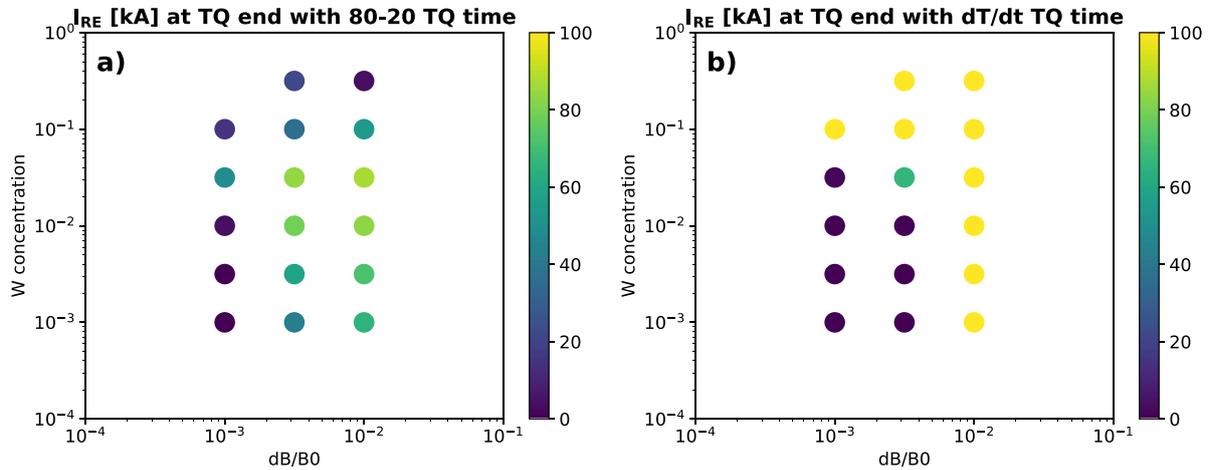

**Figure 4.** RE current at the moment of transition from TQ to CQ for fluid model simulations with different $W$ concentrations and $\delta B/B_0$ during TQ in the ASDEX-like disruption scenario. (*a*) Results with TQ time obtained from 80–20 definition. (*b*) Results from simulation with TQ time obtained from d$T$/d$t$ definition.

### 3.2. Dreicer generation

In the investigated cases for the ASDEX-like scenario, the dominant source of primary REs in the fluid model is the hot-tail mechanism. Dreicer generation can be modelled with either the Connor–Hastie formula [33], or with a neural network (NN) which was trained on results from kinetic simulations [34]. Unfortunately, the NN used currently in DREAM was trained on a number of cases which included different amounts of impurities much lighter than *W*. Despite the fact that it was proven that this NN can extrapolate to impurities not given during training, uncertainties remained about its robustness with *W* impurities. The general rule for a NN is that extreme care must be taken when using it outside of its training bounds. For this reason, we compared the NN with the Connor–Hastie model, which is known to overestimate the RE generation, so it provides an upper bound for RE generation from the Dreicer mechanism. In the isotropic simulation, the RE generation is calculated by resolving the flux of electrons through the upper boundary of the momentum grid calculated with the Fokker–Planck equation. Therefore, distinguishing between hot-tail and Dreicer RE is not possible in the isotropic case.

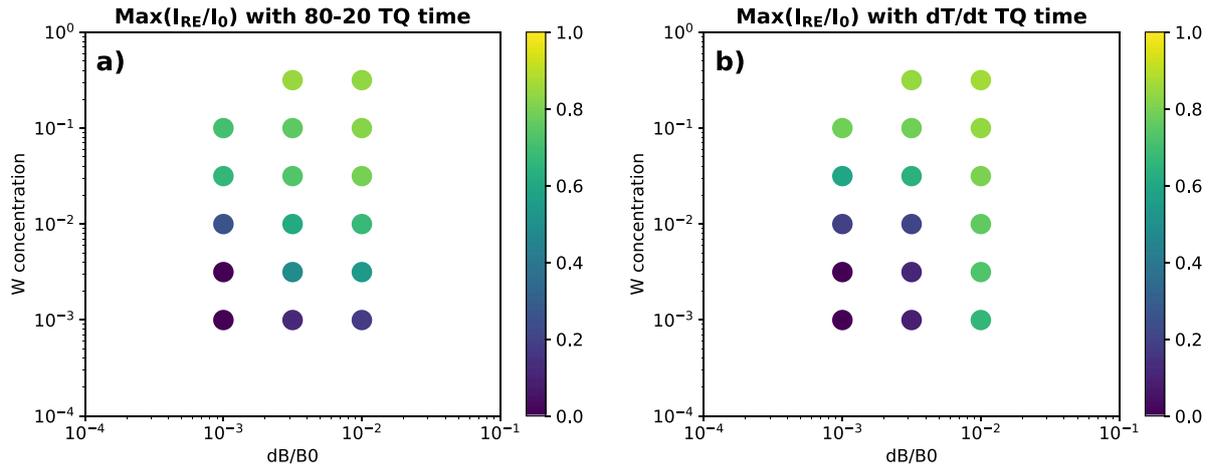

**Figure 5.** Maximum RE current compared to initial plasma current for fluid model simulations with different *W* concentrations and $\delta B/B_0$ during TQ in the ASDEX-like disruption scenario. $\delta B/B_0$ during CQ was always set to $4\times10^{-4}$. (*a*) Results with TQ time obtained from 80–20 definition. (*b*) Results from simulation with TQ time obtained from d*T*/d*t* definition.

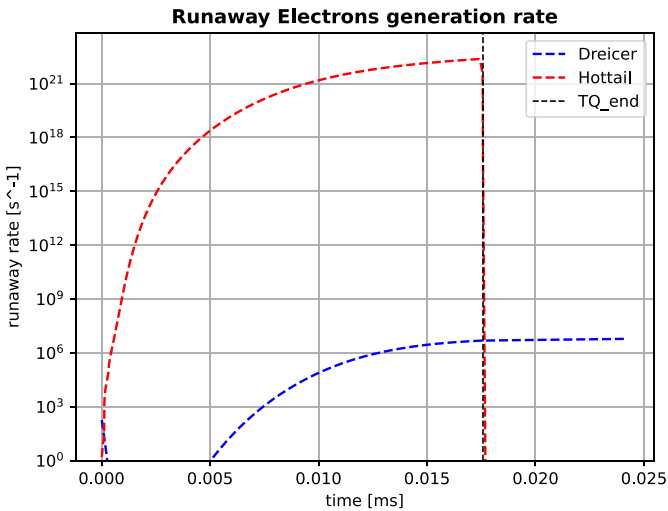

**Figure 6.** RE generation rate from Dreicer (Connor–Hastie) and hot-tail (Svenningsson) mechanisms in the ASDEX-like disruption simulation with a fluid model. Transition from TQ to CQ is marked with a dashed black line (TQ_f). *W* concentration was $1\times10^{-3}$, $\delta B/B_0$ in TQ was $1\times10^{-2}$, $\delta B/B_0$ in CQ was $4\times10^{-4}$.

The obtained results for the ASDEX-like disruption scenario show that even with the Connor–Hastie model, the Dreicer generation is several orders of magnitude smaller than hot-tail generation in every investigated case. As shown in figure 6, the strongest primary RE generation occurs during the TQ, when the enhanced transport, due to magnetic perturbations, accelerates the temperature drop and fuels the hot-tail generation mechanism.

In some cases for the ITER-like discharge, Dreicer generation was significant, as shown in figure 7. The results from some of the simulations show that the NN can give Dreicer generation values even higher than the Connor–Hastie model. It proves that in the investigated cases the NN cannot be trusted, as it can overestimate the RE generation by many orders of magnitude. For now no good solution for the fluid model exists, as Connor–Hastie is known to overestimate the Dreicer generation and the NN is unreliable with *W* impurities. Due to the unpredictable nature of errors from the machine learning algorithms, it is not recommended to use this NN in the presence of *W* impurities. The Connor–Hastie model can be used only if a high accuracy of the Dreicer generation rate is not needed. For an accurate prediction, a kinetic simulation is necessary. Retraining the NN using dedicated cases with *W* impurities is planned in the future, but it lies outside the scope of this work.

### 3.3. Fluid and Isotropic models

DREAM can be used with different levels of complexity, starting from the kinetic representation of the electron population on the momentum-space grid, through various averages and approximations up to the simple fluid-like representation. In this work, two approaches were tested. The first is the fluid model, where the bulk of electrons is assumed to have a Maxwellian distribution and the RE population is traced only in terms of its density. Primary RE generation is calculated from models of Dreicer and hot-tail generation, as described in section 3.2. The second approach is referred to as the isotropic model, where a pitch angle-averaged kinetic equation is solved. The electrons are divided into 'cold', 'hot' and RE populations. The 'cold' and RE population are calculated in almost the same way as in the fluid model. The only difference is that cold population temperature is set to very low temperature (usually 1 eV) and very low density. It later heats up by receiving electrons from the 'hot' spectrum which cool down below the momentum threshold separating both populations. The momentum grid includes electrons from the so-called 'hot' part of the momentum spectrum—their normalised momentum is well above the thermal momentum and below the specified runaway threshold, as described in [18]. This region is critical to directly simulate the RE generation rate—it is obtained from the flux of electrons through the upper boundary of the momentum grid, which replaces the fluid Dreicer and hot-tail models. The name of this approach comes from the fact that in the reduced kinetic equation, the leading order term in the expansion of the distribution function is isotropic [18]. The isotropic model is the simplest kinetic model available in DREAM. Its computation cost is roughly 3–4 times larger than that of the fluid model, which we consider a reasonable trade-off between computation time and additional insight obtained by using a more complex model.

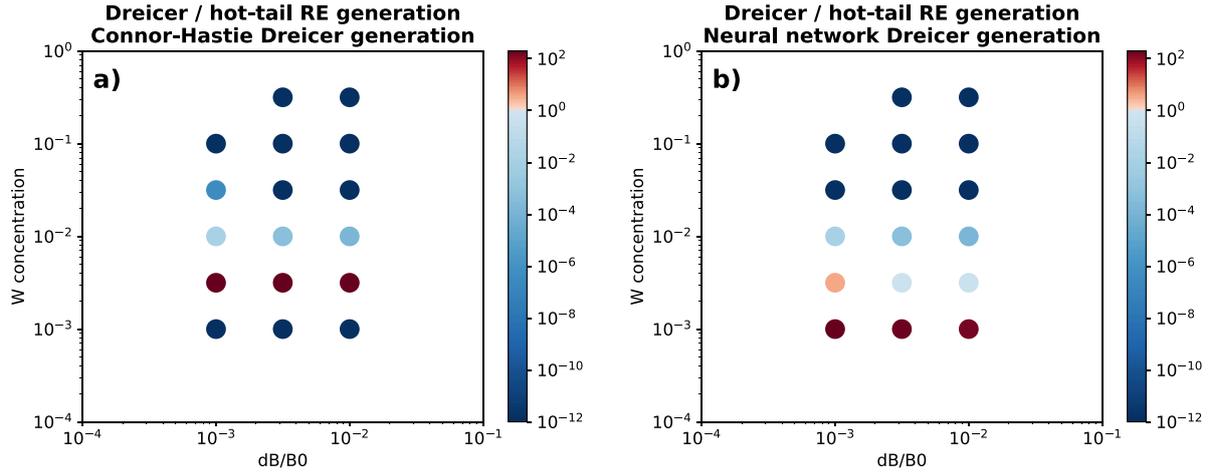

**Figure 7.** Maximum Dreicer RE generation rate compared to maximum hot-tail generation rate for simulations with different *W* concentrations and $\delta B/B_0$ during TQ. Disruption conditions were based on the ITER-like scenario, $\delta B/B_0$ during CQ was always set to $4\times 10^{-4}$. (*a*) Results from simulation with a Connor–Hastie model of Dreicer generation. (*b*) Results from simulation with NN modeling of the Dreicer generation.

The first difference between both approaches can be noticed during the TQ, after *W* impurities are ionised. In the fluid model, all non- REs are treated as one fluid, so new electrons coming from ionisation are instantly thermalized with the bulk electron population. This causes faster cooling of the initial electron population. In the isotropic model, new electrons are introduced into the 'cold' population, but the bulk population of electrons in the plasma is in the 'hot' population for the first microseconds, so the thermalization is not instantaneous. After the ionisation, the electrons from the 'hot' population will either cool down and join the 'cold' population, or accelerate and become REs—depending on their momentum. Generation of the REs is not based on the models of Dreicer and hot-tail generation, but is instead a result of kinetic equations on the momentum grid. The main results are lower RE generation and longer CQ in the isotropic model. Similar effects, but at much smaller scale, were observed in other work [31].

Results from the fluid model are more conservative than the results from the isotropic model. Fluid simulations use the Svenningsson model for hot-tail generation [35], which is not fully valid for plasma with weakly ionised impurities. It probably overestimates the RE generation, which can be the main reason for the noticeable discrepancies between both approaches, as visible in figure 8. The isotropic model, on the other hand is not well suited for the cases with low amounts of impurities, when there is a small population of 'cold' electrons compared to 'hot' electrons, as one of the assumptions of the model is that collisions amongst the 'hot' electrons are negligible. These collisions would probably decrease the RE generation and make current decay longer. Bearing in mind the differences between fluid and isotropic models, it is possible to use just the fluid model for testing RE mitigation strategies or in the initial scans of the parameter space. It can also be useful for comparison with experiments, as diagnostics for temperature and current can be used to prescribe part of the plasma evolution. The isotropic model should probably give more accurate predictions, despite the fact that collisions between hot electrons are neglected. Still, the experimental validation of these models with W impurities should be made to give a definite statement on their accuracy.

### 3.4. Impact of tokamak size

The last investigated parameter in simulations of RE generation was the tokamak size. At the same *W* concentration, the number of radiating *W* atoms scales with volume, while the heat diffused to walls comes only through the plasma surface. The radiated power should thus dominate over the transport effects in a bigger tokamak, keeping the same transport coefficients. To investigate how different ratios of these two factors will influence the disruption dynamics, we compared simulations made in two different geometries: one based on the ASDEX-like characteristics and one based on the ITER-like ones. The minor and major plasma radii are approximately 4 times bigger in the ITER-like case, see table 1. Figure 9 presents results from these simulations, which were performed with the isotropic model.

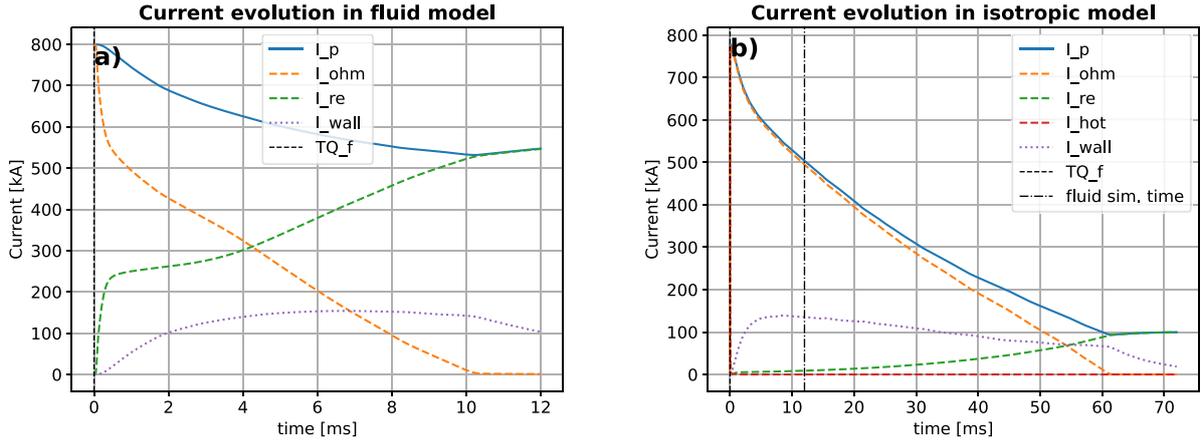

**Figure 8.** Time evolution of total plasma current (I_p), ohmic current (I_ohm), RE current (I_re), hot electrons current (I_hot), wall current (I_wall) in the ASDEX-like disruption simulation with (*a*) the fluid model and (*b*) the isotropic model of plasma. Transition from TQ to CQ is marked with a dashed black line (TQ_f). Simulation time from the fluid model case is marked in the plot (*b*) with a dash-dot line. $W$ concentration was $3.16\times10^{-2}$, $\delta B/B_0$ in TQ was $3.16\times10^{-3}$, $\delta B/B_0$ in CQ was $4\times10^{-4}$.

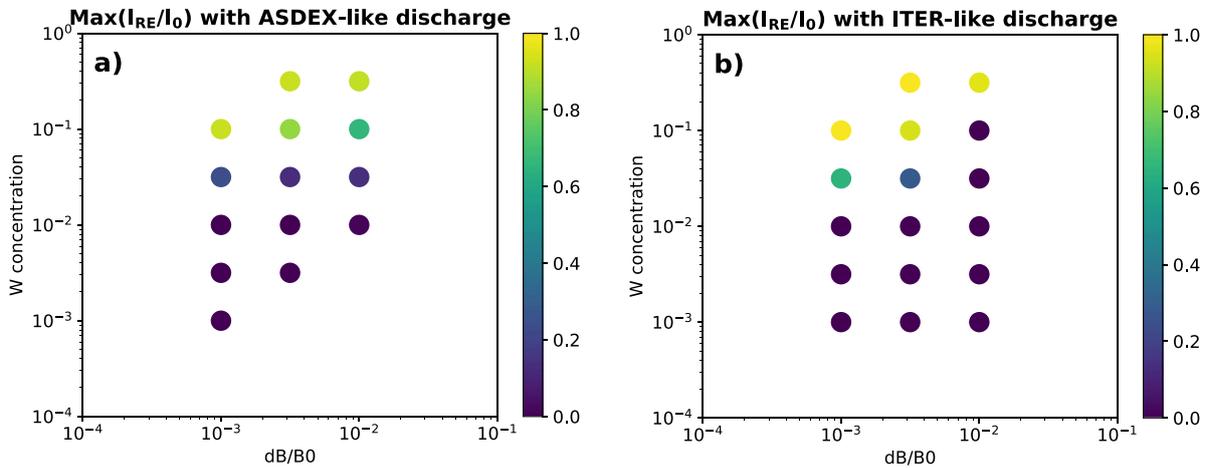

**Figure 9.** Maximum RE current compared to initial plasma current for simulations using the isotropic model with different $W$ concentrations and $\delta B/B_0$ during TQ in (*a*) ASDEX-like disruption scenario (*b*) ITER-like disruption scenario. TQ time obtained from 80–20 definition, $\delta B/B_0$ during CQ was always set to $4\times10^{-4}$.

In neither machine do plasmas with $W$ concentration up to $10^{-2}$ disrupt through a radiative collapse. Radiated power is balanced by ohmic heating, so the temperature is kept at the level of hundreds of eV for more than 0.15 s. This is a limit after which we can expect plasma disruption by vertical displacement event and excess halo current in the surrounding structures [3]. With higher tungsten concentrations, there is a negative dependence of the maximum RE current on the magnetic perturbation strength. This can be partially explained by the increased transport of the RE during TQ, which leads not only to loss of the RE, but also to their redistribution to the regions where electric field is lower and cannot sustain the avalanche growth.

There is a visible difference in maximum current conversion to RE current between both geometries. This shows that RE current on larger tokamaks could be higher than a simple scaling by maximum current would suggest. The RE current can also reach significant values at lower tungsten concentration in larger tokamaks. This is in line with the general expectations that RE generation can become more problematic when reactor-scale tokamaks are considered. Many of these issues are however related to the plasma current, so in case of compact tokamaks with high magnetic field a similar issue can occur as in ITER-size devices.

## 4. Simulation results in the ITER-like scenario

As the last step of the presented work, a series of simulations for the ITER-like disruption scenario was conducted in four different configurations. Two simulations were done with the 80–20 TQ time definition and two with the d$T$/d$t$ TQ time definition. In each case, one simulation was performed with the fluid model and one with the isotropic model of the electrons in the plasma. The resulting RE current is presented in figures 10 and 11. For ITER, the safety limit of RE current, set to prevent PFC damage, is 150 kA [23], which corresponds to 1% of the initial plasma current. In the plots, RE current which was within this limit is marked in a green colour scale and cases where the limit was breached are presented in a red colour scale.

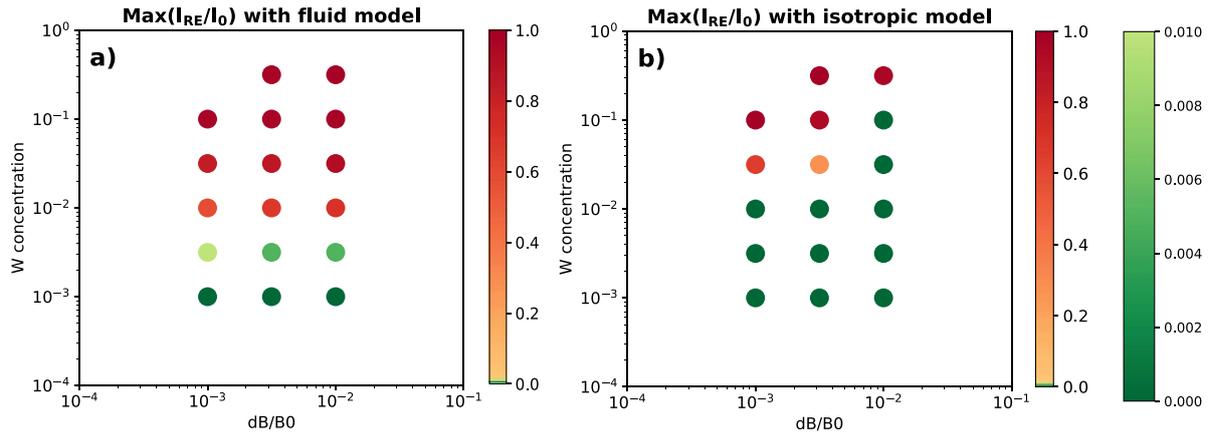

**Figure 10.** Maximum RE current compared to initial plasma current for simulations with different *W* concentrations and $\delta B/B_0$ during TQ in the ITER-like disruption scenario. TQ time obtained from the 80–20 definition. $\delta B/B_0$ during CQ was always set to $4\times10^{-4}$. (*a*) Results from simulations with the fluid model. (*b*) Results from simulations with the isotropic model.

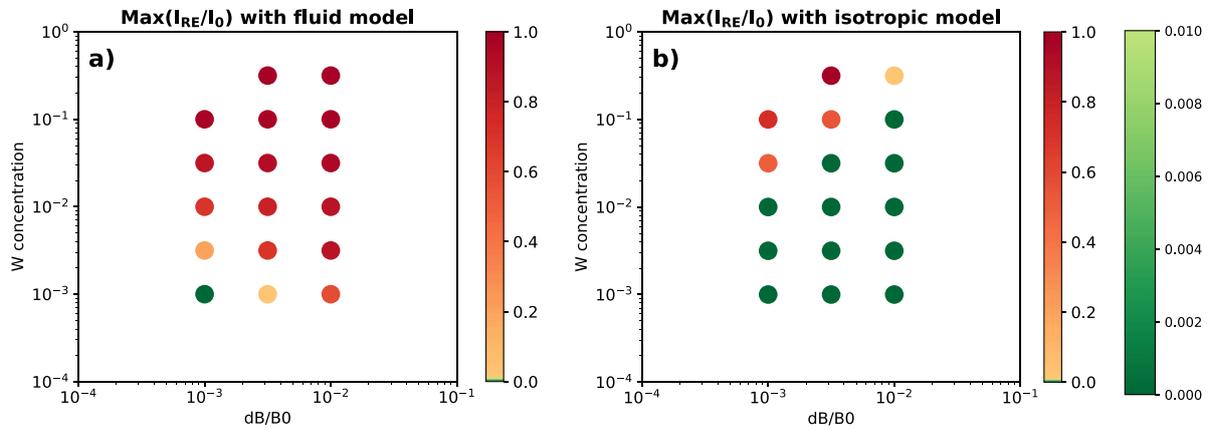

**Figure 11.** Maximum RE current compared to initial plasma current for simulations with different *W* concentrations and $\delta B/B_0$ during TQ in the ITER-like disruption scenario. TQ time obtained from the d*T*/d*t* definition. $\delta B/B_0$ during CQ was always set to $4\times10^{-4}$. (*a*) Results from simulations with the fluid model. (*b*) Results from simulations with the isotropic model.

The general result is that RE generation increases with higher *W* concentration. This seems to be a consequence of the faster plasma cooling, which enhances the hot-tail generation. In theory, an increased electron density in the plasma can increase the chance of secondary generation through avalanching. However, in our simulations, high RE currents are in some cases reached even when the *W* concentration is not high enough to significantly increase the electron density.

There is a clear discrepancy between fluid and isotropic models in terms of RE generation. The fluid model in a qualitative sense was always overpredicting the RE generation—so for safety analysis it should not give false negative answers about the risk of RE current. It is however not reliable in quantitative terms, as the exact value of RE current is, in many cases, strongly overpredicted when compared to an isotropic model. We expect the isotropic model to be more accurate than the fluid model due to more accurate calculation of the electrons dynamics in the 'hot' population, which is crucial for the detailed simulation of the RE generation. However both are just numerical models which should be verified by comparison with dedicated experiments before making any conclusive statements.

In the case of the fluid model, the magnetic perturbation usually increases the RE generation rate (the exceptions from this rule can be mostly observed when RE current is very small compared to the total plasma current). This is contrary to the results of the isotropic model. The difference is most likely due to the overestimated plasma cooling rate and resulting increased hot-tail generation in the fluid model. In general, the isotropic model should be able to describe the same physical phenomena as the fluid model, so there is no reason to assume that this is some physical phenomena not captured by one of the models. The most likely explanation is a different balance of two factors which depend on magnetic perturbation: hot-tail generation, which is increased by faster cooling, and deconfinement of RE. RE transport is the same in both models, but the isotropic model calculates the hot-tail generation with a more detailed approach, which should lead to more accurate results.

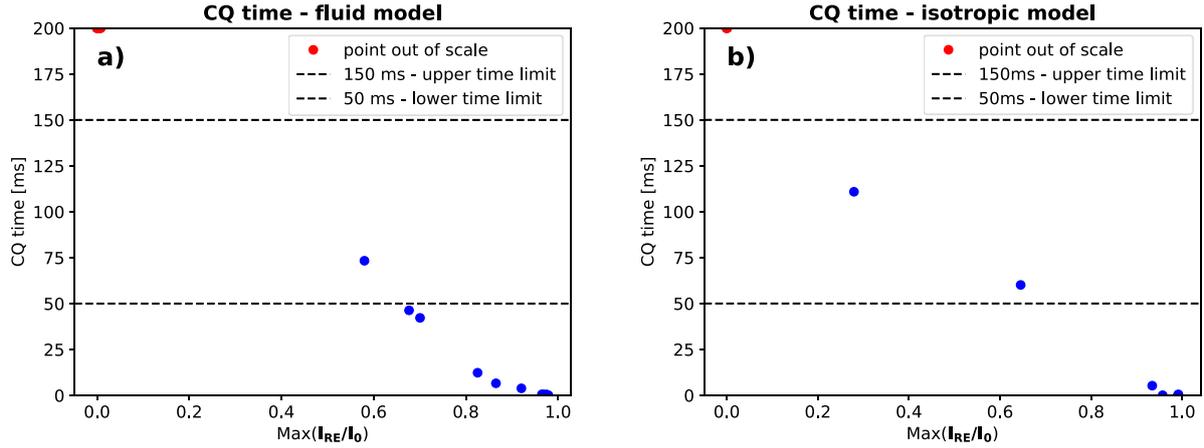

**Figure 12.** CQ time as a function of maximum ratio of RE current in the ITER-like disruption scenario. TQ time obtained from the 20–80 definition. $\delta B/B_0$ during CQ was always set to $4\times10^{-4}$. (*a*) Results from simulations with the fluid model. (*b*) Results from simulations with the isotropic model. Blue points represents CQ time for given value of max $(I_{RE}/I_0)$, red points marks simulations where CQ time was out of scale for given max $(I_{RE}/I_0)$. Horizontal lines represent minimum and maximum of allowed CQ time for ITER [3].

Figure 11 shows results of simulations with a longer TQ time, obtained from the d*T*/d*t* definition. It shows that the presented models also provide different trends related to the TQ time, as fluid simulations show in this case an extremely high RE generation risk, while isotropic simulations result in even lower RE levels than in the 80–20 TQ.

The observation of the CQ times in the simulated ITER-like disruption scenario shows that in the cases without RE generation, the CQ exceeds 0.15 s. The situation changes dramatically when a large fraction of RE current is generated, as shown on figure 12. When ohmic heating decreases, the current is transformed on the timescale of milliseconds mostly into RE current and partially into currents in the tokamak structure. Strong currents flowing in the machine structure can lead to mechanical failure due to electromagnetic forces. In the presented work, wall currents were included to account for important effects in the plasma dynamics. Analysis of the machine safety from the related forces would require separate investigations.

The CQ time is defined as:

$$\tau_{CQ} = \frac{t(I_{ohm} = 0.2 I_0) - t(I_{ohm} = 0.8 I_0)}{0.6} \quad (2)$$

where $I_{ohm}$ is the ohmic component of the plasma current and $I_0$ is the plasma current at the beginning of the simulation (pre-disruption current).

## 5. Outlook and summary

Our results show that RE generation caused by tungsten impurities may become a significant problem only at tungsten concentrations above $10^{-3}$, which is unlikely to occur during normal tokamak operation. As long as an abnormal event does not introduce large amounts of tungsten dust or flakes into the plasma, W-induced RE should not pose a direct threat to the tokamak operation.

The RE current depends on the TQ dynamics, which is different than in the case of intentional disruption mitigation by massive material injection. The exact TQ time and magnetic perturbation are difficult to predict due to uncertainties in the initial conditions and high computational cost, though a worst case scenario can be defined for safety analysis. For such application, the fluid model can be sufficient, but a number of limitations connected to this approach should be kept in mind. The isotropic model, which is the simplest kinetic model available in DREAM, can be used at a computation cost roughly 3–4 times larger than the fluid model [18], providing additional insight into electron dynamics. To create a truly self-consistent simulation, it would be necessary to include the magnetic surface evolution, especially magnetic surface stochastization and healing. This however was outside the scope of the presented work, as it would require a different modelling approach and significantly more computational resources.

Results from the fluid model are more conservative than the results from the isotropic model. This comes from higher RE generation rates, which are probably overestimated. The isotropic model, on the other hand, is not well suited for the cases with low amounts of impurities, where electrons coming from introduced impurities do not dominate over the electrons already present in plasma. However, the aspects neglected in the isotropic model should only overestimate the results, so there is a solid basis to assume that the results are still conservative. Bearing in mind the differences between fluid and isotropic models, it is possible to use only the fluid model for testing RE mitigation strategies or in the initial scans of the parameter space, while the isotropic model should be used to assess the accuracy of the solution. A more definitive statement about accuracy would require dedicated experiments for validation of the models.

The TQ time cannot be accurately predicted without MHD simulations. When using approximated TQ time definitions, a conservative approach should be used which, on the one hand, ensures a significant temperature drop (so the radiation is stronger, or at least balances the ohmic heating) and on the other hand the initial RE seed should not be deconfined by the magnetic perturbation. Magnetic perturbation strength can have a significant effect on the RE generation in medium size tokamaks, but for ITER the exact value is less important for RE generating scenarios, as radiation losses are dominant in these cases.

In the analysed cases, the hot-tail RE generation mechanism was dominant as a primary source of RE. The Dreicer generation mechanism cannot be reliably estimated using the NN proposed by Hesslow *et al* [34]. When verification with kinetic simulation is not available, it is recommended to use the Connor–Hastie model, as it sets an upper bound of the Dreicer generation providing conservative, but not necessarily accurate, estimations of the RE generation. The NN should be retrained on the cases which include tungsten impurities before it can be used in simulation tools.

Simulations performed for the ITER-like scenario are moderately concerning. The first analysis with the fluid model suggests that RE generation is possible if tungsten intrusion causes a rise of impurity concentration to a value of $10^{-3}$ or higher. This would be concerning if left alone, as such levels can be reached in tokamaks [17, 36, 37]. However, our analysis made with the isotropic model shows that concentrations of tungsten on a level of 1% or lower should not lead to a significant RE current generation on its own. As mentioned before, the isotropic model should be more trustworthy. A 3 mm droplet of tungsten evaporated into the core plasma would be required to reach a tungsten concentration of 1%, which is unlikely. It should be kept in mind that in the case of additional RE sources, tungsten will probably increase the generation rate. When the conditions for avalanche become favourable, the transition to runaway current is fast and can lead to a catastrophic discharge with a few MA runaway current. In such cases, the CQ time would probably exceed the safety limits predicted for ITER, causing a risk of mechanical damage of the vacuum vessel due to electromagnetic forces from currents induced in the structure.

An open question is the impact of tungsten impurities on the planned mitigation system for ITER. Mitigation with shattered pellet injection requires balance between RE suppression and ohmic current preservation (to avoid a too fast CQ) [38]. This creates an operational space for the mitigation system in terms of the neon and deuterium quantities that should be injected into the plasma. With the presence of tungsten impurities, this operational space can become narrower or shifted in relation to pure D-T plasma without impurities. This requires further investigations which are outside the scope of this work. Furthermore, estimation of tungsten ablation and evaporation in the burning plasma conditions can be different than in present-day tokamaks, as the high-energy particles present in such plasmas will be able to penetrate the ablated sheet of material that normally shields the impurities and reduce the evaporation rate. This can lead to an unprecedented rate of tungsten concentration, which would be detrimental for the tokamak integrity.

We believe that the presented results of disruption simulations with high tungsten content are just a first step and emphasise the need for further studies on this topic. The current simulations require validation efforts, as tungsten-induced disruptions were not initiated on purpose in current tokamaks, so existing data is scarce and fragmented. Dedicated research on this topic should help to evaluate the risk posed by tungsten impurities and its effects on the mitigation systems of reactor-scale tokamaks.

## Acknowledgments

The authors would like to thank T. Fülöp and I. Pusztai for the fruitful discussions and support. Also they would like to thank Martin O'Mullane and Thomas Pütterich for their advice on the selection of ADAS datasets and their comments about tungsten specific issues.

J. Walkowiak acknowledges financial support provided by the Polish National Agency for Academic Exchange NAWA under the Programme STER—Internationalisation of doctoral schools, Project no. PPI/STE/2020/1/00020". We gratefully acknowledge Poland's high-performance computing infrastructure PLGrid (HPC Centers: ACK Cyfronet AGH) for providing computer facilities and support within computational Grant No. PLG/2022/015994. This work has been carried out within the framework of the EUROfusion Consortium, funded by the European Union via the Euratom Research and Training Programme (Grant Agreement No. 101052200—EUROfusion). Views and opinions expressed are however those of the author(s) only and do not necessarily reflect those of the European Union or the European Commission. Neither the European Union nor the European Commission can be held responsible for them. This project is co-financed by the Polish Ministry of Education and Science in the framework of the International Co-financed Projects (PMW) programme Contract No. 5450/HEU—EURATOM/2023/2.

## Appendix A. ADAS data selection

Most of the atomic data in DREAM is retrieved from the ADAS database [25] through the OPEN-ADAS system from the ADF11 class. The datasets used in the code are: data for ionisation (SCD), recombination (ACD), line power (PLT) and recombination/bremsstrahlung power (PRB). Charge-exchange cross-coupling coefficients (CCD) are not used in DREAM, but were implemented to maintain compatibility with the rest of the data, because the same datasets are used also by the related code STREAM [39]. For CCD there is only one dataset available for every element, so the selection is straightforward. For the remaining parameters, there are few datasets to choose from. These datasets are marked by numbers which are sometimes referred to as years, although in the case of tungsten this is just a jargon, as these numbers do not correspond to the years of creation of the dataset in any way. The selection of the most suitable dataset is usually done by referring to the ADAS recommendation (https://open.adas.ac.uk/man/appxa-11.pdf), which unfortunately does not include tungsten. The selection of the most advanced datasets for tungsten was therefore done after consultation with the authors of the ADAS database. The datasets were obtained from a spectrum of raw atomic data of different quality. Selected databases were created with a larger number of atomic states included and more detailed approximations of wave functions representing atomic states, so they are the most accurate estimations currently available. For SCD, ACD and PRB files, the year 50 was chosen while for the PLT file, the year 42 was selected.

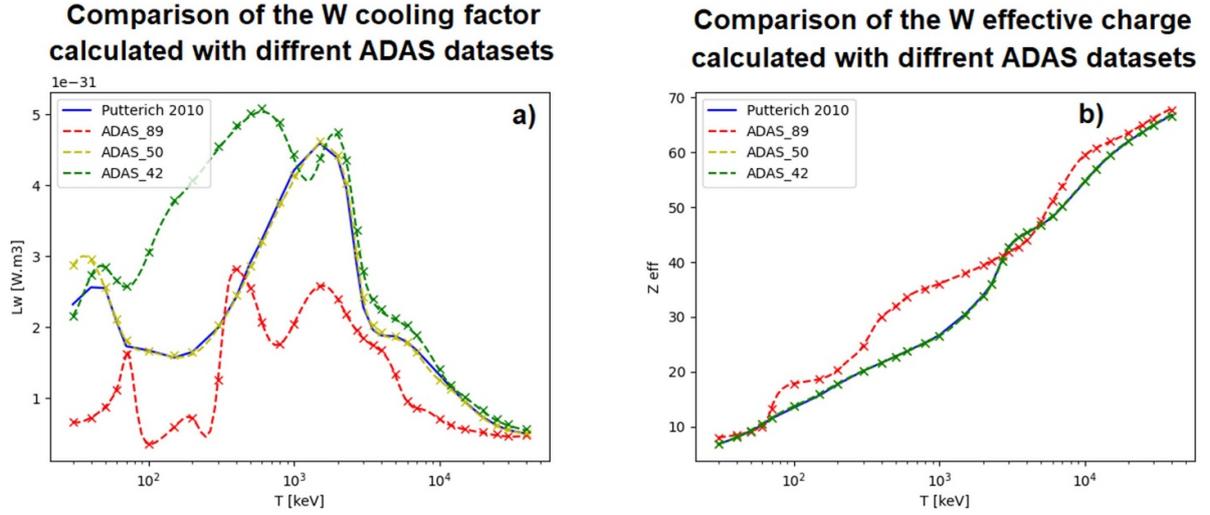

**Figure 13.** Comparison of (*a*) the *W* cooling factor and (*b*) the effective charge of *W* ions as a function of the electron temperature, obtained with different versions of the ADAS datasets. For ADAS_89 and ADAS_50, all datasets were from the year 89 and 50, respectively. In case of ADAS_42, the PLT file is from year 42 and SCD, ACD, PRB files are taken from the year 50.

It was found during the validation of the code that for low temperatures, the cooling factor calculated with the PLT year 50 dataset was different from the corresponding data published in Putterich [40]. The difference is smaller than the uncertainty in the atomic data for this temperature range, so it is not considered to be problematic, but the exact source of the difference could not be concluded. For completeness, we present in figure 13 the comparison of the results from different datasets.

## Appendix B. MEE

The MEE of *W* ions was calculated using the modified LPA as described in [27, 41]:

$$\mathrm{MEE} = \exp\left[\frac{4\pi}{N}\int r_i^2 \rho(r_i) \ln(\hbar\omega_0(r_i))\,\mathrm{d}r_i + F\right] \quad (3)$$

where *N* is the number of electrons bound to the considered ion, $r_i$ is the radial coordinate of the ion, $\rho(r_i)$ is the electron density distribution, which is assumed to be spherically symmetric around the nucleus, $\omega_0(r_i)$ is the local plasma frequency of the electron gas in the ion [42] and *F* is a correction factor derived by fitting the LPA to the results from Sauer *et al* [28, 29]:

$$F = \exp\left(\frac{Z-N}{Z}\right) - 0.9 \quad (4)$$

where *Z* is the atomic number of the ion.

The electron density distribution $\rho(r_i)$ was calculated with the so-called optimised Pratt–Tseng (PT$_{\mathrm{opt}}$) model as described in [41] The expression for $\rho(r_i)$ is:

$$\rho(r_i) = \frac{1}{4\pi r_i}\left[\sum_{s=1}^{5}\frac{N_s}{a_s^2}\exp\left(-\frac{r_i}{a_s}\right)\right] \quad (5)$$

where $N = \sum_{s=1}^{5} N_s$ is the number of electrons bound to the ion, $N_s$ is the number of electrons included in each summation part and $a_s$ are parameters corresponding to each group of electrons.

The grouping of electrons is given in table 2.

The $a_s$ coefficients can be approximated with the following equations:

$$a_s(Z, N) = 1\bigg/\sqrt{\lambda_s^2\frac{(1-x^{n_s+1})}{1-x}} \quad (6)$$

$$\lambda_s(Z) = c_{1,s}Z^{c_{2,s}} \quad (6.1)$$

$$n_s(Z) = c_{3,s}Z^{c_{4,s}} \quad (6.2)$$

where $x = \frac{Z-N}{Z}$ and *Z* is the atomic number of the considered ion. The optimised $c_{n,s}$ parameters are presented in table 3.

**Table 2.** Grouping of electrons in the PT$_{opt}$ model Adapted from [41]. CC BY 4.0.

| Electron group | $N_1$ | $N_2$ | $N_3$ | $N_4$ | $N_5$ |
|---|---|---|---|---|---|
| Max. number of bound electrons in each group | 2 | 8 | 18 | 28 | Rest |
| Total bound electrons when group fully occupied | 2 | 10 | 28 | 54 | Rest |

**Table 3.** Optimised parameters for PT$_{opt}$ model Adapted from [41]. CC BY 4.0.

| | | $i=1$ | $i=2$ | $i=3$ | $i=4$ | $i=5$ |
|---|---|---|---|---|---|---|
| $\lambda_i(Z)$ | $c_{1,s}$ | 1.1831 | 0.1738 | 0.0913 | 0.0182 | 0.7702 |
| | $c_{2,s}$ | 0.8368 | 1.0987 | 0.9642 | 1.2535 | 0.2618 |
| $n_{s,i}(Z)$ | $c_{3,s}$ | 0.3841 | 0.6170 | 1.0000 | 1.0000 | 1.0000 |
| | $c_{4,s}$ | 0.5883 | 0.0461 | 1.0000 | 1.0000 | 1.0000 |

## ORCID iDs

J. Walkowiak ● https://orcid.org/0000-0002-9787-1691
M. Hoppe ● https://orcid.org/0000-0003-3994-8977
I. Ekmark ● https://orcid.org/0000-0001-8065-4650
A. Jardin ● https://orcid.org/0000-0003-4910-1470
J. Bielecki ● https://orcid.org/0000-0002-3460-8677
Y. Savoye-Peysson ● https://orcid.org/0000-0001-8594-9474
D. Mazon ● https://orcid.org/0000-0001-5560-2277
D. Dworak ● https://orcid.org/0000-0002-0768-1748
M. Scholz ● https://orcid.org/0000-0002-7330-1782